\documentclass[lettersize,journal]{IEEEtran}
\usepackage{amsmath,amsfonts}
\usepackage{algorithmicx}
\usepackage{algorithm}
\usepackage{algpseudocode}
\usepackage{array}
\usepackage[caption=false,font=normalsize,labelfont=sf,textfont=sf]{subfig}
\usepackage{textcomp}
\usepackage{stfloats}
\usepackage{url}
\usepackage{verbatim}
\usepackage{graphicx}
\usepackage{cite}
\usepackage{soul}
\usepackage{amsmath,amsfonts}
\usepackage{algorithmicx}
\usepackage{algorithm}
\usepackage{algpseudocode}
\usepackage{booktabs}
\usepackage{multirow}
\usepackage{xcolor}
\usepackage{pifont}
\usepackage{tablefootnote}

\usepackage[colorlinks=true,urlcolor=black,citecolor=black,linkcolor=black]{hyperref}

\begin{document}

\title{On the Adversarial Robustness of Learning-based Image Compression Against Rate-Distortion Attacks}

\author{Chenhao Wu, Qingbo Wu,~\IEEEmembership{Member,~IEEE,} Haoran Wei, Shuai Chen, Lei Wang, King Ngi Ngan,~\IEEEmembership{Life Fellow,~IEEE,} Fanman Meng,~\IEEEmembership{Member,~IEEE,} Hongliang Li,~\IEEEmembership{Senior Member,~IEEE}
\thanks{The Authors are with the School of Information and Communication Engineering, University of Electronic Science and Technology of China, Chengdu, 611731, China (e-mail: chwu@std.uestc.edu.cn; qbwu@uestc.edu.cn; hrwei@std.uestc.edu.cn; schen@std.uestc.edu.cn, lwang@std.uestc.edu.cn; knngan@uestc.edu.cn; hlli@uestc.edu.cn; fmmeng@uestc.edu.cn)}
\thanks{Corresponding author: Qingbo Wu.}}



\maketitle

\begin{abstract}
Despite demonstrating superior rate-distortion (RD) performance, learning-based image compression (LIC) algorithms have been found to be vulnerable to malicious perturbations in recent studies. However, the adversarial attacks considered in existing literature remain divergent from real-world scenarios, both in terms of the attack direction and bitrate. Additionally, existing methods focus solely on empirical observations of the model vulnerability, neglecting to identify the origin of it. These limitations hinder the comprehensive investigation and in-depth understanding of the adversarial robustness of LIC algorithms. To address the aforementioned issues, this paper considers the arbitrary nature of the attack direction and the uncontrollable compression ratio faced by adversaries, and presents two practical rate-distortion attack paradigms, \textit{i.e.}, Specific-ratio Rate-Distortion Attack (SRDA) and Agnostic-ratio Rate-Distortion Attack (ARDA). Using the performance variations as indicators, we evaluate the adversarial robustness of eight predominant LIC algorithms against diverse attacks. Furthermore, we propose two novel analytical tools for in-depth analysis, \textit{i.e.}, Entropy Causal Intervention and Layer-wise Distance Magnify Ratio, and reveal that \textit{hyperprior} significantly increases the bitrate and \textit{Inverse Generalized Divisive Normalization (IGDN)} significantly amplifies input perturbations when under attack. Lastly, we examine the efficacy of adversarial training and introduce the use of online updating for defense. By comparing their advantages and disadvantages, we provide a reference for constructing more robust LIC algorithms against the rate-distortion attacks.
\end{abstract}

\begin{IEEEkeywords}
Robust image compression, rate-distortion attack, adversarial examples
\end{IEEEkeywords}

\section{Introduction}
\IEEEPARstart{I}{mage} compression is a fundamental task in the field of image processing and has been extensively studied over the past few decades. Generations of traditional handcrafted codecs have facilitated the image storage and transmission, \textit{e.g.}, JPEG \cite{wallace1992jpeg}, HEVC Intra \cite{sullivan2012overview}, and VVC Intra \cite{bross2021overview}. In recent years, there has been a growing interest within the image processing community towards learning-based image compression (LIC), which achieves superior rate-distortion (RD) performance with end-to-end optimized models \cite{minnen2018joint,Cheng_2020_CVPR,zou2022devil,liu2023tcm,koyuncu2024efficient,shi2023rate,hong2020efficient,ge2024nlic,zhang2024machine,wang2022deep}.

Although learning-based image compression methods have demonstrated significant potential in practical applications within real-world scenarios, it is essential for them to undergo rigorous verification on the robustness. One of the most severe challenges in this regard is the security against adversarial attacks, commonly known as adversarial robustness. This means defending against the malicious samples generated from adversaries who attack the compression system \cite{carlini2019evaluating}. 

\begin{figure}
	\centering
	\includegraphics[width=\linewidth]{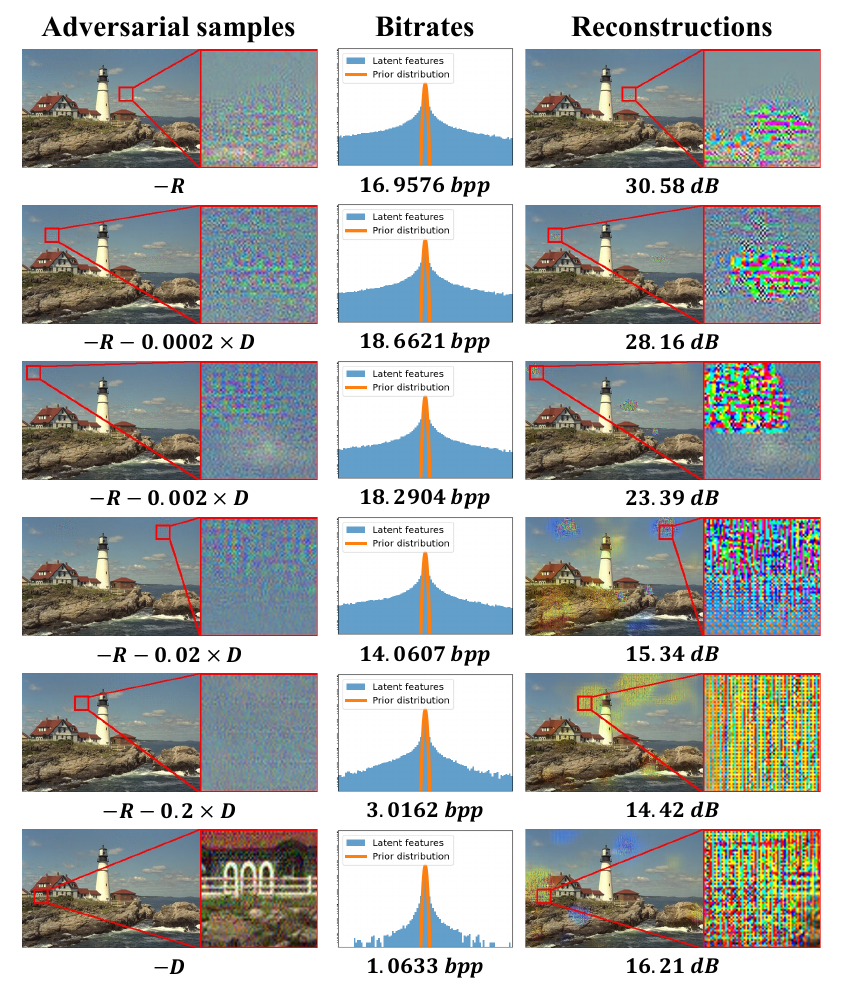}
	\caption{Visualization of input images and compression results of \textit{Minnen2018} under different attack directions. The middle column visualizes the difference between the distribution of latent representation and the prior distribution for lossless entropy coding.}
	\label{fig:rda_samples}
\end{figure}

Recently, several works \cite{chen2023towards,sui2023reconstruction,liu2023manipulation,cao2024enhancing,zhu2024attack,song2024training} have explored the vulnerability of the learning-based image compression to adversarial attacks. By applying malicious perturbations to input image, adversaries can induce the compression model to produce corrupted reconstruction \cite{chen2023towards,sui2023reconstruction,cao2024enhancing,zhu2024attack,song2024training} or significant increase in bitrate \cite{liu2023manipulation,cao2024enhancing,zhu2024attack,song2024training}. However, practical attacks are not limited to these two directions, but can be any joint direction of bitrate attack and distortion attack. As shown in Fig. \ref{fig:rda_samples}, the attack results produced by these joint attacks exhibit significant diversity. Notably, some joint attacks can even inflict greater damage compared to single-dimensional attacks focusing solely on either bitrate or distortion. This indicates that existing studies are insufficient in fully examining the robustness of LIC algorithms. Additionally, the compression ratio is usually agnostic to users in many real-world application, which is dynamically adjusted with the storage space and the communication bandwidth. In this case, the adversarial robustness of the LIC algorithm will exhibit quite different properties.

\begin{figure}
	\centering
	\includegraphics[width=\linewidth]{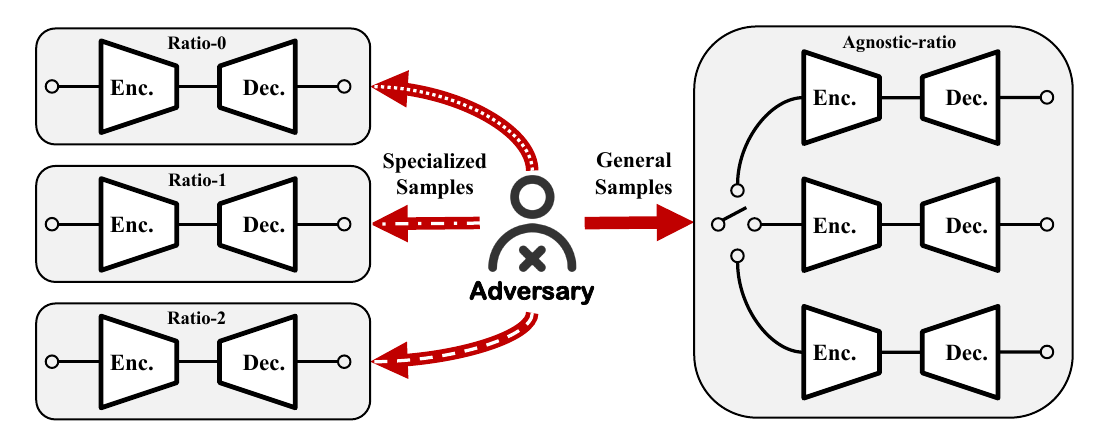}
	\caption{\textbf{Left:} Specific-ratio Rate-Distortion Attack (SRDA), where adversarial samples are designed for specific compression ratios. \textbf{Right:} Agnostic-ratio Rate-Distortion Attack (ARDA), which seeks to attack all compression ratios with general adversarial samples.}
	\label{fig:head}
\end{figure}

To simulate these practical attacks, we introduce two novel attack paradigms, \textit{i.e.}, the Specific-ratio Rate-Distortion Attack (SRDA) and the Agnostic-ratio Rate-Distortion Attack (ARDA), as shown in Fig. \ref{fig:head}. Specifically, SRDA generates dedicated adversarial samples for specific compression ratios, while ARDA tries to destruct the entire compression algorithm at all compression ratios with a single malicious image. In both paradigms, adversaries maintain the flexibility to degrade the compression performance in arbitrary directions within the RD plane. In the scope of this paper, we regard a LIC method as a integrated algorithm, and the models trained for different rate-distortion trade-offs as its submodels.

Besides, existing researches are limited to empirical observations of the vulnerability of the LIC algorithm, but lacks identification on the origin of it. Inspired by the causal intervention in explaining adversarial attacks \cite{ren2022towards,tang2021adversarial} and the latent responses \cite{leeb2022exploring}, we propose two novel analytical tools, \textit{i.e.}, Entropy Causal Intervention (ECI) and Layer-wise Distance Magnify Ratio (LDMR). Specifically, ECI locates the model vulnerabilities exploited by bitrate attacks through investigating the impact on the final bitrate of each component in the entropy model. While LDMR locates the model vulnerabilities exploited by distortion attacks through quantitatively calculating the amplification of disturbances by each network layer. Using these two tools, we reveal the significant vulnerabilities in two commonly used modules \textit{hyperprior} and \textit{IGDN} for the first time.

On the basis of attack paradigms and analytical tools, we investigate the efficacy of adversarial training in defensing against rate-distortion attacks. We also adopt the online updating technology in adversarial defense for the first time. By comparing the improvements on RD performance of the two defense methods, we summarize their applicable scenarios, advantages, and disadvantages. This proveds a reference for the construction of more adversarially robust LIC algorithms in the future.

In summary, contributions of this paper are as follows:
\begin{itemize}
	\item Two novel attack paradigms are proposed to simulate the diverse attacks in real-world applications, covering multiple possible attack directions and attack objects. Two novel analytical tools are introduced to identify the orgin of vulnerabilities in LIC models.
	\item Extensive experiments on eight predominant LIC algorithms are conducted to evaluate the adversarial robustness against rate-distortion attacks. The significant role of \textit{hyperprior} and \textit{IGDN} in the vulnerability of LIC models is revealed for the first time.
	\item The efficacy of adversarial training and online updating against rate-distortion attacks is comprehensively examined and compared, providing a reference for the development of more adversarially robust LIC algorithms.
\end{itemize}

\section{Related Work}
\label{sec:related}
\subsection{Learning-based Image Compression}
With the rapid development of deep learning techniques, a multitude of learning-based image compression approaches have demonstrated superior rate-distortion performance compared to conventional codecs\cite{minnen2018joint,Cheng_2020_CVPR,zou2022devil,liu2023tcm,koyuncu2024efficient,shi2023rate,hong2020efficient,ge2024nlic,zhang2024machine,wang2022deep}. The prevailing image compression methods are constructed on autoencoder architecture and entropy models. Prototype models of the prevailing autoencoder-type image compression were initially proposed and developed in \cite{balle2017end} and \cite{ballevariational}, with the factorized prior model and hyperprior model still playing crucial roles in state-of-the-art approaches. 

In \cite{minnen2018joint} and \cite{Cheng_2020_CVPR}, spatial context information was extracted from the decoded latent features through masked convolutional layers to improve the modeling of prior distribution. Later, \cite{minnen2020channel} developed a channel context entropy model to leverage the dependencies between channels. To improve the spatial modeling capability, \cite{zou2022devil} and \cite{liu2023tcm} adopted the structure of transformer to better capture the local and global information. 

In contrast to the separately trained submodels for each bitrate in these algorithms, \cite{song2021variable} and \cite{cai2022high} adopted the variable-rate design. By applying different transformations on a set of basic parameters, they acheved multiple compression ratios with a single compression model.

\subsection{Adversarial Attack}
Deep learning models have been observed to be vulnerable under adversarial attacks \cite{szegedy2014intriguing}, which are typically executed by introducing imperceptible disturbances to the regular input. \cite{szegedy2014intriguing} agrued the inscrutable model results come from the discontinuity of input-output mappings to a significant extend, and generated adversarial examples through box-constrained L-BFGS algorithm. \cite{goodfellow2014explaining} claimed the vulnerability comes from the linear mapping, and proposed a fast gradient sign method (FGSM) to generate disturbances, which was developed to iterative versions I-FGSM \cite{kurakin2016adversarial} and PGD \cite{madry2017towards}. \cite{moosavi2015deepfool} accurately evaluated the robustness of models through comparing the minimal perturbations that can change the output of networks. \cite{carlini2017towards} developed a powerful attack method by solving a constrained optimization problem.

Recently, the vulnerability of individual submodels in LIC algorithms has been investigated in several studies \cite{chen2023towards,sui2023reconstruction,liu2023manipulation,cao2024enhancing,zhu2024attack,song2024training}. To assess the distortion vulnerability, \cite{chen2023towards} developed a fast threshold-constrained distortion attack method, while \cite{sui2023reconstruction} mitigated the perceptibility of the malicious disturbance by imposing constraints on low frequences. To examine the bitrate vulnerability, \cite{liu2023manipulation} raised the bitrate using malicious perturbations while maintaining reconstruction quality. \cite{zhu2024attack} investigated the effects of FGSM \cite{goodfellow2014explaining} and PGD \cite{madry2017towards} on bitrate attack and distortion attack. \cite{cao2024enhancing} experimented the effect of bitrate attack and distortion attack under the constraint of constant distortion and bitrate, respectively.

In \cite{yu2023backdoor}, researchers explored the vulnerability of LIC algorithms under the setting of backdoor attacks. It is worth noting that \cite{yu2023backdoor} aims to change the parameters of the image encoder to form a backdoor that can be attack by pre-defined triggers. Differently, our work does not change the model parameters, but investigates the inherent adversarial robustness of the LIC algorithms by applying malicious perturbations on the input image.

Effect of joint rate-distortion attacks on learning-based video compression models have been studied in \cite{chang2022rovisq}. It focused on the Group of Pictures (GOP) in temporal coding, and emphasized on the empirical observation of the adversarial robustness of three video codecs. In contrast, our paper concentrates on image compression models, and conducts extensive experiments on eight LIC methods that cover multiple types of network structures. Besides, we construct analytical tools for in-depth analysis, and identify the origin of vulnerability in LIC models.

\begin{figure}
	\centering
	\includegraphics[width=0.75\linewidth]{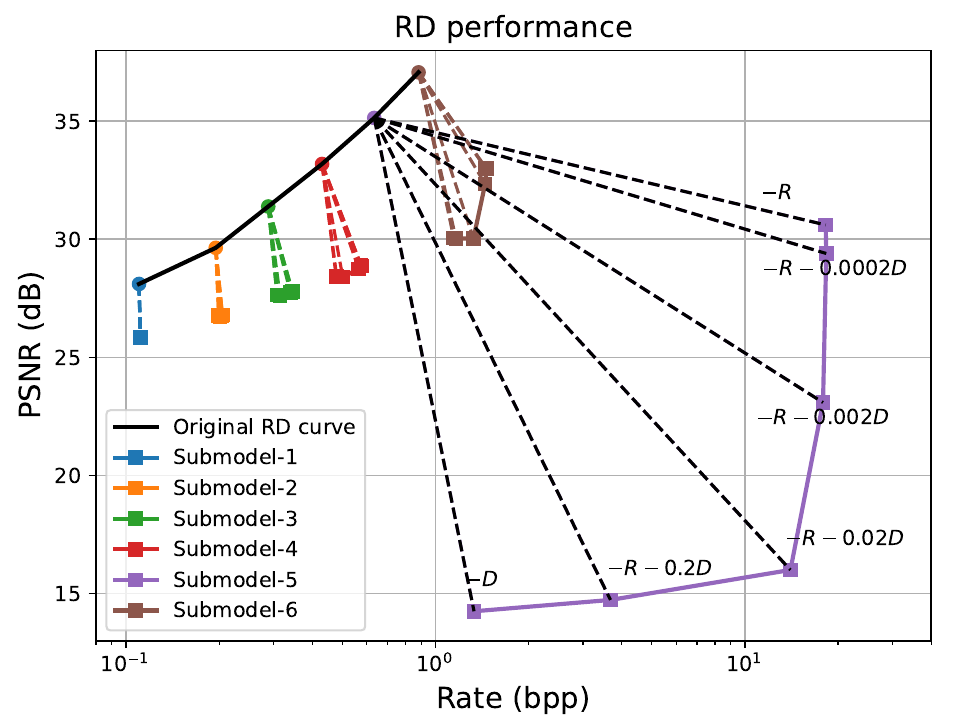}
	\caption{RD performance of Minnen2018 on adversarial images specifically designed for submodel-5.}
	\label{fig:rda_performance}
\end{figure}
\section{Preliminary}
Learning-based image compression model usually contains an image encoder $g_a(\cdot)$ for feature extraction and an image decoder $g_s(\cdot)$ for reconstruction, together with an entropy model for losslessly encoding quantized features. When compressing, the input image $x$ is first transformed to latent representation $y=g_a(x)$, and then quantized to $\hat{y}=Q(y)$ to execute entropy encoding. Inversely, $\hat{y}$ is entropy decoded and transformed back to reconstructed image $\hat{x}=g_s(\hat{y})$. During training, the loss function is usually defined as a trade-off between the bit-rate $\mathcal{R}(x)$ consumed by compressed binary strings and the distortion $\mathcal{D}(x)$ in decoded images, which can be formulated as the rate-distortion loss:
\begin{equation}
	\label{equ:rd}
	\mathcal{L}_{rd} = \mathcal{R}(x) + \lambda \mathcal{D}(x),
\end{equation}
where $\lambda$ is the trade-off factor that controls the balance between two loss items.

Adversarial examples are often defined as inputs that are designed to cause the model to make mistakes \cite{xu2020adversarial}. Given an input image $x$, it can be modified to $x_a$ by adding malicious noise $\delta$:
\begin{equation}
	\label{equ:adv_image}
	x_a = x + \delta, \quad \lVert \delta \rVert \leq \epsilon, \quad x_a \in [0,1].
\end{equation}
$\epsilon$ is the maximum level of disturbance, $\lVert \cdot \rVert$ is defined as the averaged $l_2$-norm in this paper. By intentionally designing the perturbation $\delta$, it is expected that the RD performance on the input $x_a$ will exhibit significant degradation compared to the original input $x$. Since most image compression algorithms used in practical applications are standardized, researches on adversarial robustness are often conducted within the context of white-box attacks \cite{chen2023towards,sui2023reconstruction}. 

\section{Attack Paradigms and Assessment}

\subsection{Specific-ratio Rate-Distortion Attack}

\begin{algorithm}[t]
	\renewcommand{\algorithmicrequire}{\textbf{Input:}}
	\renewcommand{\algorithmicensure}{\textbf{Output:}}
	\caption{Specific-ratio Rate-Distortion Attack (SRDA)}
	\label{alg:srda}
	\begin{algorithmic}[1]
		\Require Image $x \in \mathbb{R}^{H \times W \times C}$, attack coefficients $\gamma_r$ and $\gamma_d$, maximum disturbance $\epsilon$, attack steps $\mathcal{T}$ on the surface of $\epsilon$-sphere.
		\State \textbf{Initialization:} $\delta = \vec{0}^{H \times W \times C}$
		\Repeat
		\While{$\lVert \delta \rVert>\epsilon$}
		\State ${\arg \min} _\delta \lVert \delta \rVert$
		\EndWhile
		\State Calculate $\mathcal{L}_s$ based on $\gamma_r$ and $\gamma_d$. \Comment{Eq. (\ref{equ:srda})}
		\State Update $\delta$ based on $\partial \mathcal{L}_s / \partial \delta$.
		\Until{$\mathcal{T}$ attacks are made after the first reach of the $\epsilon$-sphere surface}
		\Ensure $\max(\min(x_a, 1), 0)$
	\end{algorithmic}
\end{algorithm}

When the compression ratio to be used by the target LIC algorithm can be determined by the adversaries, the specially crafted adversarial samples can exert maximum effect. In response to this situation, we developed the Specific-ratio Rate-Distortion Attack (SRDA) to carry out attacks in various directions, as shown in Algorithm \ref{alg:srda}. The perturbation $\delta$ can be inferred by optimizing Eq. (\ref{equ:srda}) under the constraint of Eq. (\ref{equ:adv_image}):
\begin{equation}
	\label{equ:srda}
	\begin{aligned}
		\mathop{\arg\min}\limits_{\delta} \, \mathcal{L}_s = - \gamma_r \cdot \mathcal{R}(x_a) - \gamma_d \cdot \mathcal{D}(x_a),
	\end{aligned}
\end{equation}
where $\gamma_r$ and $\gamma_d$ are attack coefficients controlled by adversaries. When $\gamma_r$ and $\gamma_d$ are set to $0$ and $1$, adversarial samples are crafted to increase reconstruction distortion, as considered in \cite{chen2023towards} and \cite{sui2023reconstruction}. When $\gamma_r$ is set to $1$ and $\gamma_d$ is set to a negative value, it turns into the bitrate attack without compromising reconstruction quality in \cite{liu2023manipulation}.

We primarily investigate six attack directions, where $(\gamma_r,\gamma_d)$ are set to $(1,0)$, $(1,0.0002)$, $(1,0.002)$, $(1,0.02)$, $(1,0.2)$, and $(0,1)$, respectively. Given the high correlation among similar attack directions, we can utilize the results from these six directions to estimate unexplored performance. The implementation of SRDA follows a similar practice to \cite{chen2023towards}, the difference is that we control the number of attack steps after the perturbation reaching the surface of $\epsilon$-sphere for fairness. Take \textit{submodel-5} of Minnen2018 \cite{minnen2018joint} as an example, we show the RD performance before and after SRDA in Fig. \ref{fig:rda_performance} purple lines. The significant variation in RD performance across different attack directions is evident.

\subsection{Agnostic-ratio Rate-Distortion Attack}

\begin{algorithm}[t]
	\renewcommand{\algorithmicrequire}{\textbf{Input:}}
	\renewcommand{\algorithmicensure}{\textbf{Output:}}
	\caption{Agnostic-ratio Rate-Distortion Attack (ARDA)}
	\label{alg:frda}
	\begin{algorithmic}[1]
		\Require Image $x \in \mathbb{R}^{H \times W \times C}$, attack coefficients $\gamma_r$ and $\gamma_d$, maximum disturbance $\epsilon$, attack steps $\mathcal{T}$ on the surface of $\epsilon$-sphere.
		\State \textbf{Initialization:} $\delta = \vec{0}^{H \times W \times C}$
		\Repeat
		\While{$\lVert \delta \rVert>\epsilon$}
		\State ${\arg \min} _\delta \lVert \delta \rVert$
		\EndWhile
		\For{$k \in [1,N]$}
		\State Calculate $\mathcal{L}_{s}^{(k)}$ based on $\gamma_r$ and $\gamma_d$. \Comment{Eq. (3)}
		\State Calculate $w^{(k)}$. \Comment{Eq. (5)}
		\EndFor
		\State Calculate $\mathcal{L}_f$ based on $\mathcal{L}_{s}^{(k)}$ and $w^{(k)}$. \Comment{Eq. (4)}
		\State Update $\delta$ based on $\partial \mathcal{L}_f / \partial \delta$.
		\Until{$\mathcal{T}$ attacks are made after the first reach of the $\epsilon$-sphere surface}
		\Ensure $\max(\min(x_a, 1), 0)$
	\end{algorithmic}
\end{algorithm}

However, when the compression ratio to be used by the target LIC algorithm can not be determined by the adversaries, the efficacy of specially crafted adversarial samples can not be guaranteed. As shown in Fig. \ref{fig:rda_performance}, the efficacy of adversarial samples specially designed for \textit{submodel-5} exhibits a notable decrease when compressed by other submodels.

To ensure the effectiveness of malicious attacks, adversaries need to design adversarial samples targeting a broader range of bitrates. Consequently, we develop the Agnostic-ratio Rate-Distortion Attack (ARDA), as shown in Algorithm \ref{alg:frda}. The loss function of it can be extended from SRDA as follows:
\begin{equation}
	\label{equ:frda}
	\begin{aligned}
		\mathop{\arg\min}\limits_{\delta} \, \mathcal{L}_f &= \sum_{k=1}^{N} w^{(k)} \mathcal{L}_{s}^{(k)},
	\end{aligned}
\end{equation}
where $N$ is the number of submodels to attack, $k$ indicates the index of submodel, $\mathcal{L}_{s}^{(k)}$ represents the $\mathcal{L}_s$ of SRDA on submodel $k$, $w^{(k)}$ is the weight used to balance the attack effect on submodel-$k$ with $\sum_{k=1}^{N} w^{(k)} = 1$.

Given the inconsistent adversarial robustness among submodels, employing identical weights could lead to the concentration of attack impacts on specific bitrates. To mitigate this issue and distribute the attack effects evenly across all bitrates, we dynamically adjust the weights based on the observed attack effects throughout the attack procedure:
\begin{equation}
	\label{equ:weighted}
	w^{(k)} = \frac{\exp(w'^{(k)} / \tau)}{\sum_i \exp(w'^{(i)} / \tau)}, \quad
	w'^{(k)} = \mathcal{L}_{s-init}^{(k)} / \mathcal{L}_{s}^{(k)},
\end{equation}
where $\tau$ is the temperature coefficient that controls the difference among $w^{(k)}$. 

\subsection{Metrics and Tools}
\label{sec:tools}
We introduce the \textit{Performance Variation} to assess the adversarial robustness of LIC methods. To identify the origin of vulnerability, we propose two analytical tools, \textit{i.e.}, \textit{Entropy Causal Intervention} and \textit{Layer-wise Distance Magnify Ratio}.

\subsubsection{Performance Variation}
Based on the compression results on the benign images and the maliciously perturbed images, the variation in RD performance can be formulated as:
\begin{equation}
	\label{eq:global}
	\Delta \mathcal{R} = \mathcal{R}(x_a) - \mathcal{R}(x), \quad \Delta \mathcal{D} = \mathcal{D}(x_a) - \mathcal{D}(x).
\end{equation}
Through these global metrics, the adversarial robustness of LIC models can be assessed quantitatively. In practice, we use the mean ($\mu(\Delta \mathcal{R})$, $\mu(\Delta \mathcal{D})$) and standard deviation ($\sigma(\Delta \mathcal{R})$, $\sigma(\Delta \mathcal{D})$) calculated over all the submodels and/or attack directions to compare LIC algorithms from various perspectives.

For a fine-grained analysis, we further localize the performance variations into patch-wise bitrate analysis and pixel-wise distortion analysis:
\begin{equation}
	\label{eq:local}
	\Delta_s \mathcal{R} = \mathcal{R}_s(y_a) - \mathcal{R}_s(y), \quad \Delta_s \mathcal{D} = \mathcal{D}_s(x_a) - \mathcal{D}_s(x),
\end{equation}
where the subscript $s$ indicates that the spatial information is retained during the calculation of RD performance. Through these local indicators, we are able to visualize and understand the characteristic of LIC methods under adversarial attacks.

\begin{figure}[t]
	\centering
	\includegraphics[width=0.7\linewidth]{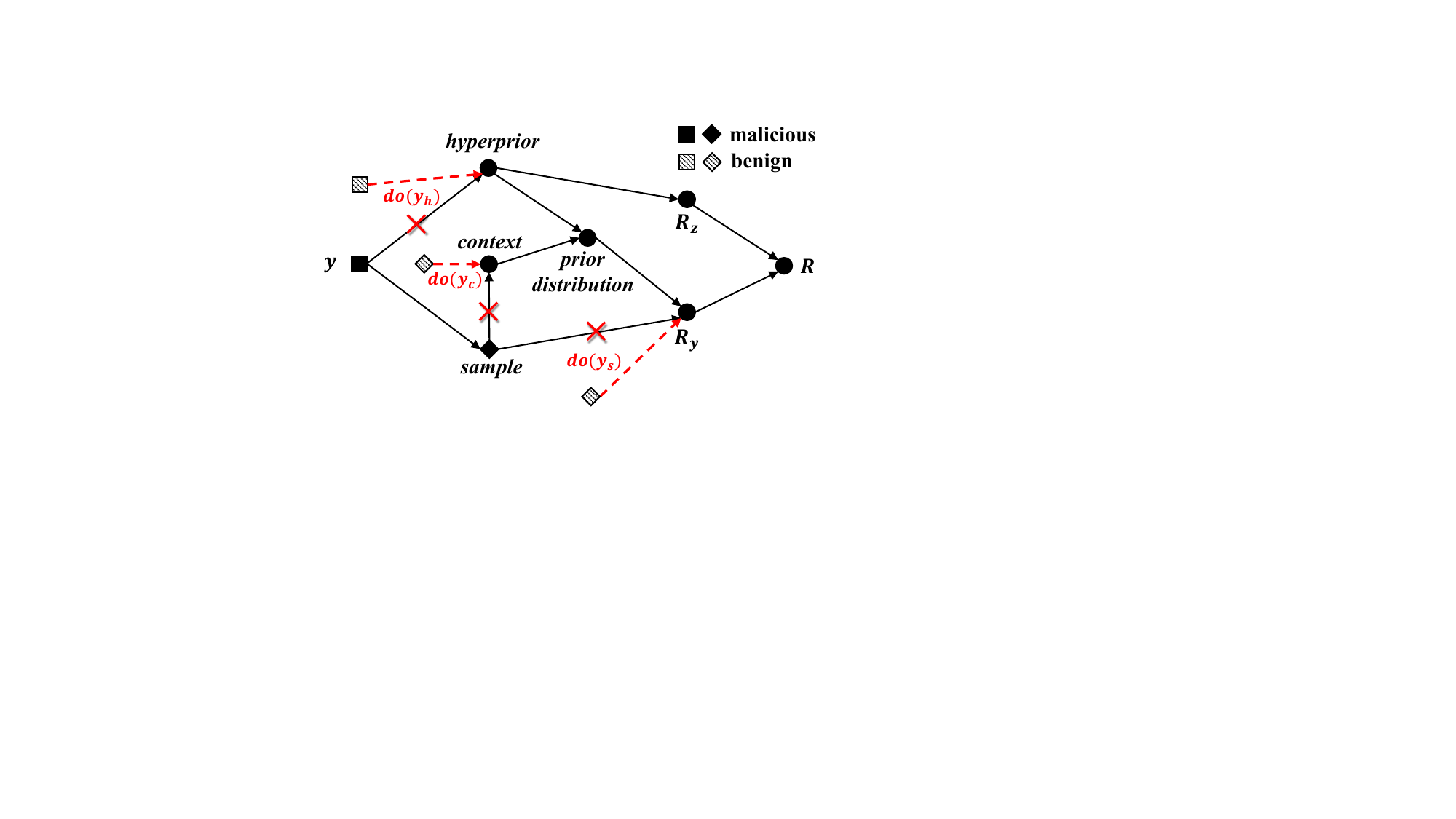}
	\caption{Causal diagram of entropy model in LIC methods. Do-operators denote replace malicious features with benign features, and are highlighted in red.}
	\label{fig:causal}
\end{figure}
\subsubsection{Entropy Causal Intervention}
According to the principle of arithmetic coding \cite{langdon1984introduction,witten1987arithmetic}, the compressed bitrate is determined by both the encoding samples and the prior distribution shared by both encoder and decoder. In most LIC algorithms, this distribution is estimated from decoded samples (\textit{context}) and global side information (\textit{hyperprior}). To understand the model vulnerability related to entropy coding, we follow the concepts in causal analysis \cite{ren2022towards,tang2021adversarial} and construct the causal diagram of entropy model in Fig. \ref{fig:causal}. With the help of do-operators, the vulnerability of each branch can be assessed through the causal variability of bitrate $\mathcal{R}$. Specifically, $\textbf{do}(y_s)$, $\textbf{do}(y_c)$, and $\textbf{do}(y_h)$ represent replace malicious features of samples, context features, and hyperprior features with benign counterparts, respectively.

\begin{figure}[t]
	\centering
	\includegraphics[width=\linewidth]{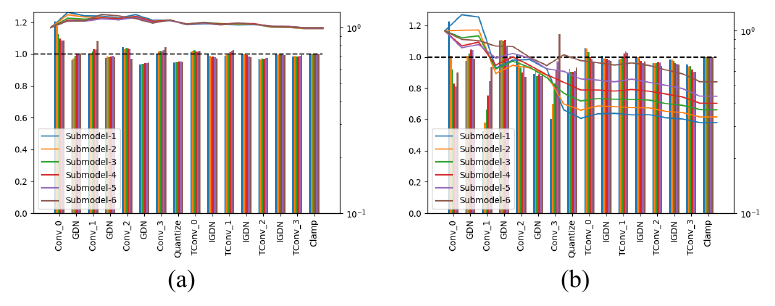}
	\caption{(a) LDMR and CDMR calculated on two benign images, as well as (b) on a benign input image and its gaussian noisy version. The horizontal axis arranges the network layers from left to right in order. The bar graph corresponds to the left coordinate, representing the value of LDMR, and the line graph corresponds to the right coordinate, representing the value of CDMR.}
	\label{fig:LDMR}
\end{figure}
\subsubsection{Layer-wise Distance Magnify Ratio}
Among the components of LIC methods, the distortion is determined by image encoder $g_a(\cdot)$, quantizer $Q(\cdot)$, and image decoder $g_s(\cdot)$. For simplicity, we define these distortion related layers as a function $f$, and refer the $i$-th layer as $f_i$. Layers before $f_i$ is represented as $f_{[0,i-1]}$, layers include and after $f_i$ is represented as $f_{[i,I]}$, where $I$ is the index of the last model layer. Then we can write $f$ as $f_{[i,I]} \circ f_{[0,i-1]}$, where $\circ$ denotes the composite function.

Given a benign input image $x$ and its reconstructed version $f(x)$ after encoding and decoding, we have:
\begin{equation}
	\label{eq:recon}
	x \approx f(x) = f_{[i,I]} \circ f_{[0,i-1]}(x),
\end{equation}
then we can induce:
\begin{equation}
	\label{eq:recompress}
	\begin{aligned}
		f_{[0,i-1]}(x) &\approx f_{[0,i-1]} \circ f(x) \\
		&= f_{[0,i-1]} \circ f_{[i,I]} \circ f_{[0,i-1]}(x).
	\end{aligned}
\end{equation}

However, this approximate equality no longer holds on the maliciously perturbed image $x_a=x+\delta$. Considering that the reconstruction $f(x_a)$ does not contain malicious disturbances, the first $f_{[0,i-1]}$ in the right term of Eq. (\ref{eq:recompress}) is not affected by adversarial attacks. Consequently, the difference between left and right items is mainly due to the effect of malicious perturbations on $f_{[i,I]}$.

Movited by this, we propose a novel indicator named Layer-wise Distance Magnify Ratio (LDMR) to reflect the amplify ratio of each network layer on disturbance:
\begin{equation}
	\label{eq:ldmr_a}
	LDMR_{[i,I]} = \frac{\lVert f_{[0,i-1]} \circ f(x_a) - f_{[0,i-1]} \circ f(x) \rVert _1}{\lVert f_{[0,i-1]}(x_a) - f_{[0,i-1]}(x) \rVert _1},
\end{equation}
\begin{equation}
	LDMR_i = \frac{LDMR_{[i,I]}}{LDMR_{[i+1,I]}}.
\end{equation}
To reflect the cumulative amplification effect after $i$-th network layer on the distance, we calculate the Cumulated Distance Magnify Ratio as following:
\begin{equation}
	\centering
	CDMR_i = LDMR_{[0,i]} = \prod_{k=0}^{k=i} LDMR_i.
\end{equation}

The reconstruction error between $x$ and $f(x)$ in Eq. (\ref{eq:recon}) and Eq. (\ref{eq:recompress}) comes from the compression information loss. Although the subtraction in Eq. (\ref{eq:ldmr_a}) includes the elimination of it, there still exists a certain degree of error due to the difference between compressing $x_a$ and $x$. To verify the influence of this error, we conduct two experiments on benign images in Fig. \ref{fig:LDMR}. Specifically, Fig. \ref{fig:LDMR}(a) gives the LDMR and CDMR on the distance between two benign images with distinct contents. It is evident that the indicators are around $1$, indicating that the compression algorithm has almost no effect on the distance between them. Fig. \ref{fig:LDMR}(b) shows the LDMR and CDMR on the distance between a benign input image and its gaussian noisy version. It is found that the final CDMR and the LDMR of most layers are below $1$. This phenomenon is consistent with the expectation, since the intrinsic characteristic of compression algorithm is to discard unimportant information. We can also observe that the final CDMR increases with the increase of bitrate, which reflects the fact that submodels at low bitrate will discard more trivial noise. In this context, the abnormal amplification of LDMR and CDMR caused by malicious disturbance is very obvious.

\section{Defense Techniques}

\subsection{Adversarial Training}
\label{sec:AT}
Adversarial training (AT) is one of the most effective defense methods against attacks of known types \cite{bai2021recent}. Its fundamental concept is to improve the adversarial robustness of learning-based models by training them with adversarial samples. Consequently, the essence of this method lies in achieving comprehensive coverage of the adversarial example space. We conduct experiments to investigate the efficacy of AT in defensing against joint rate-distortion attacks.

The adversarial samples generated by rate-distortion attacks are diverse, which is constrained by $\gamma_r$, $\gamma_d$, and $w^{(k)}$. To mitigate the influence of $w^{(k)}$, we conduct the adversarial training on malicious samples generated by SRDA. To effectively counter attacks from arbitrary directions and encompass a wide array of malicious images, we generate adversarial samples by randomly adjusting attack coefficients $\gamma_r$ and $\gamma_d$. 

\subsection{Online Updating}
\label{sec:online}
Online updating, initially introduced in \cite{campos2019content}, stands as a content-adaptive optimization technology. Due to the common practice of training LIC algorithms on a vast dataset of images, the resulting model often yields compromised RD performance across all training images. Online updating iteratively updates the latent representations \cite{campos2019content,yang2020improving} or parameters of encoder \cite{lu2020content} for a specific input, aimed at ehancing the RD performance. This process does not require additional training, and the decoding procedure remains unchanged.

To make the online updating in the adversarial defense closer to the attack process, we directly update the input image $x_a$:
\begin{equation}
	\label{equ:online}
	\begin{aligned}
		\mathop{\arg\min}\limits_{x_a} \, \mathcal{L}_{online} = \mathcal{R}(x_a) + \lambda \cdot \mathcal{D}(x_a),
	\end{aligned}
\end{equation}
where $\lambda$ denotes the same trade-off factor with training. Although the defender cannot optimize in the exact reversed direction of the attack due to the unknown attack coefficients $\gamma_r$ and $\gamma_d$ in Eq. (\ref{equ:srda}), this formula still can improve the RD performance towards higher compression efficiency.

\section{Experiments and Findings}
\label{sec:expr}
\subsection{Experimental Settings}
\subsubsection{Models}
We conduct experiments on Ball\'e2016\textsuperscript{\ref{compressai}} \cite{balle2017end}, Ball\'e2018\textsuperscript{\ref{compressai}} \cite{ballevariational}, Minnen2018\footnote{\href{https://github.com/InterDigitalInc/CompressAI}{https://github.com/InterDigitalInc/CompressAI}\label{compressai}} \cite{minnen2018joint}, Cheng2020\textsuperscript{\ref{compressai}} (self-attention variant) \cite{Cheng_2020_CVPR}, Song2021\footnote{\href{https://github.com/micmic123/QmapCompression}{https://github.com/micmic123/QmapCompression}} \cite{song2021variable}, Cai2022\footnote{\href{https://github.com/CaiShilv/HiFi-VRIC}{https://github.com/CaiShilv/HiFi-VRIC}} \cite{cai2022high}, Zou2022\footnote{\href{https://github.com/Googolxx/STF}{https://github.com/Googolxx/STF}} \cite{zou2022devil} and Liu2023\footnote{\href{https://github.com/jmliu206/LIC_TCM}{https://github.com/jmliu206/LIC\_TCM}} \cite{liu2023tcm} based on CompressAI platform \cite{begaint2020compressai}. These methods encompass various important structures in the field of image compression and are highly representative. Key components of each method are shown in Tab. \ref{label:components}. A submodel with fixed compression ratio is designed for each bitrate in separately trained methods. For a fair comparison, 6 submodels under $1.0$ \textit{bpp} are selected for the evaluation of robustness against ARDA. In variable-rate methods, a set of base parameters is shared by all bitrates. In order to compare with separately trained algorithms, we set the quality level of variable-rate methods to $10,25,40,55,70,90$ as the six submodels.

\begin{table*}[t]
	\centering
	\caption{Key components of victim models. Conv., Res., Attn., Den., and Trans. denote convolutional layer, residual block, attention block, dense block, and swin transformer block, respectively. SFT, INN, and LRP represent Spatial Feature Transform, Invertible Neural Network, and Latent Residual Prediction, respectively. (I)GDN stands for Generalized Divisive Normalization (GDN) and Inverse Generalized Divisive Normalization (IGDN).}
	\label{label:components}
	\resizebox{0.95\linewidth}{!}{
		\begin{tabular}{lcccc} \toprule
			& Structure Type & Transform Coding Layers & Entropy Mdoel & Method Type \\ \hline
			Ball\'e2016 \cite{balle2017end} & CNN-based & Conv. + (I)GDN & Factorized & Separately trained \\
			Ball\'e2018 \cite{ballevariational} & CNN-based & Conv. + (I)GDN & Hyperprior & Separately trained \\
			Minnen2018 \cite{minnen2018joint} & CNN-based & Conv. + (I)GDN & Hyperprior \& Spatial context & Separately trained \\
			Cheng2020 \cite{Cheng_2020_CVPR} & CNN-based & Conv. + Res. + Attn. + (I)GDN & Hyperprior \& Spatial context & Separately trained \\
			Song2021 \cite{song2021variable} & CNN-based & Conv. + Res. + SFT + (I)GDN & Hyperprior & Variable-rate \\
			Cai2022 \cite{cai2022high} & Mixed INN-CNN & Conv. + Den. + INN + SFT & Hyperprior \& Spatial context & Variable-rate \\
			Zou2022 \cite{zou2022devil} & Transformer-based\tablefootnote{Convolutional layers are used only in the input and output layers, while the transform coding network is mainly composed of transformer blocks. We follow \cite{zou2022devil,liu2023tcm} and refer to it as a transformer-based method.} & Conv. + Trans. + LRP & Hyperprior \& Channel context & Separately trained \\
			Liu2023 \cite{liu2023tcm} & Mixed Transformer-CNN & Conv. + Res. + Trans. + LRP + (I)GDN & Hyperprior \& Channel context & Separately trained \\ \bottomrule
		\end{tabular}
	}
\end{table*}

\subsubsection{Datasets}
Evaluation of the attack and defense is performed on Kodak dataset \cite{TrueColo62:online}, which contains 24 full color images with resolution $512 \times 768$. Adversarial training is performed on 14500 pictures with resolution around $1024 \times 1024$ from the web. The training images are preprocessed to remove undesirable artifacts following \cite{ballevariational}.

\subsubsection{Attack}
All the attack is conducted with \textit{Adam} optimizer with learning rate of $10^{-2}$, which is decreased to $10^{-3}$ when attacking $\mathcal{T}/2$ steps on the surface of $\epsilon$-sphere. The total step of adversarial attack $\mathcal{T}$ is set to $64$ by default, $\epsilon$ is set to $10^{-3}$ in all experiments. Temperature coefficient $\tau$ in the weight of ARDA is set equal to $\gamma$.

\subsubsection{Defense}
In adversarial training, training images are randomly cropped to $256 \times 256$. On the basis of normally trained models, the compression models to be protected is optimized on both normal and perturbed images by \textit{Adam} for $1000$ iterations. The learning rate is initialized to $10^{-4}$ and decreased to $10^{-5}$ during the last $10\%$ of training steps. In online updating, the learning rate is set to $10^{-2}$.

\subsection{Performance Variations Under SRDA}
\begin{figure*}[t]
	\centering
	\includegraphics[width=0.8\linewidth]{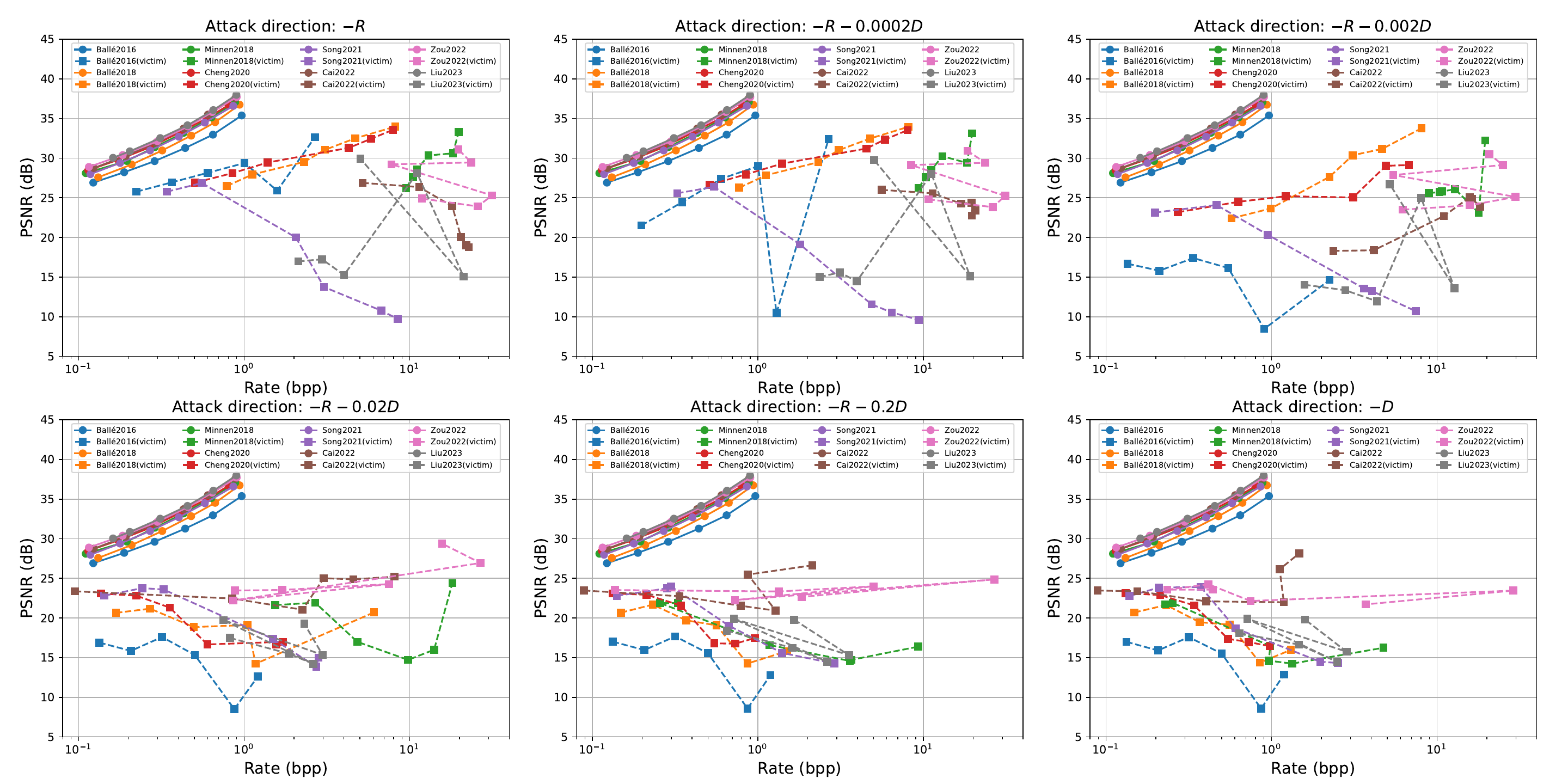}
	\caption{RD performance under six attack directions of SRDA. The performance of the same compression algorithm before and after an attack is depicted using solid and dashed lines in the same color.}
	\label{fig:rd_curves}
\end{figure*}

To understand the damage SRDA inflicts to the compression performance, we present the RD curves before and after adversarial attack within a plot. As shown in Fig. \ref{fig:rd_curves}, each plot reflects the attack effect from a direction. It can be observed that the adversarial robustness at each bitrate of the same algorithm exhibits similar patterns, but there is no strict regularity between them. For instance, all submodels of Zou2022 demonstrate inferior robustness in bitrate directions and superior robustness in distortion directions among eight LIC algorithms. However, the dotted lines between these submodels are irregular and messy. To conduct a comprehensive assessment, we evaluate eight LIC algorithms from the perspective of overall robustness, robustness in different attack directions, and robustness of submodels.

\begin{table}
	\centering
	\caption{Averaged performance variations $(\downarrow)$ over all submodels and attack directions. The {\color{blue}best} and {\color{red}worst} adversarial robustness among all methods are highlighted in blue and red.}
	\label{tab:overall}
	\resizebox{0.8\linewidth}{!}{
		\begin{tabular}{l|rr|rr} \toprule
			\multicolumn{1}{c|}{\multirow{2}{*}{Methods}} & \multicolumn{4}{c}{SRDA} \\ \cline{2-5}
			& $\mu (\Delta \mathcal{R})$ & $\mu (\Delta \mathcal{D})$ & $\sigma (\Delta \mathcal{R})$ & $\sigma (\Delta \mathcal{D})$ \\ \hline
			Ball\'e2016 \cite{balle2017end} & {\color{blue}0.2974} & {\color{red}2158.45} & {\color{blue}0.3051} & {\color{red}2240.76} \\
			Ball\'e2018 \cite{ballevariational} & 1.6635 & 521.69 & 1.4792 & 379.95 \\
			Minnen2018 \cite{minnen2018joint} & 8.5247 & 649.08 & 3.5128 & 403.64 \\
			Cheng2020 \cite{Cheng_2020_CVPR} & 1.5294 & 434.28 & 1.2946 & 268.29 \\
			Song2021 \cite{song2021variable} & 1.8357 & 1796.32 & 1.7483 & 1712.94 \\
			Cai2022 \cite{cai2022high} & 7.6381 & 310.75 & 3.3936 & 178.19 \\
			Zou2022 \cite{zou2022devil} & {\color{red}12.4988} & {\color{blue}180.11} & {\color{red}9.0529} & {\color{blue}72.33} \\
			Liu2023 \cite{liu2023tcm} & 3.9357 & 1426.99 & 3.0803 & 839.32 \\ \bottomrule
		\end{tabular}
	}
\end{table}

\begin{figure}
	\centering
	\includegraphics[width=\linewidth]{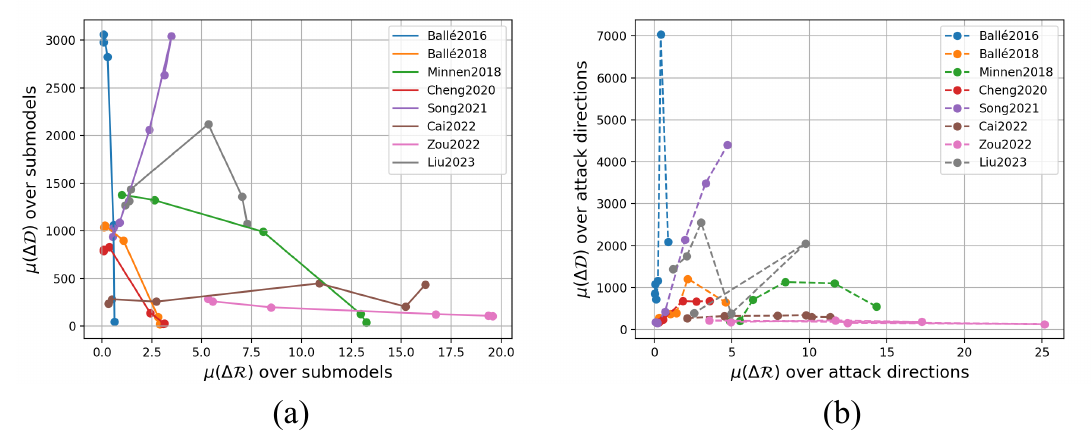}
	\caption{(a) Averaged performance variations over all submodels and (b) averaged performance variations over all attack directions of SRDA.}
	\label{fig:delta_SRDA}
\end{figure}

\subsubsection{Overall Robustness}
Based on the compression results before and after malicious attacks, we calculate the mean and standard deviation of performance variations over all the submodels and attack directions. As shown in Table \ref{tab:overall}, bitrate and distortion of Ball\'e2016 exhibit the best and worst adversarial robustness among all the compression methods, respectively. On the contrary, Zou2022 demonstrates the worst bitrate robustness and the best distortion robustness against malicious attacks. Overall, Cheng2020 exhibits the best adversarial robustness among these algorithms. 

It is worth noting that large performance variations comes with large standard deviation, indicating a big difference between the robustness of submodels and different attack directions. This makes it necessary to conduct further analysis from the perspective of attack direction and submodel.

\subsubsection{Robustness in Different Attack Directions}
Averaged performance variations $\mu(\Delta \mathcal{R})$ and $\mu(\Delta \mathcal{D})$ calculated over submodels can reflect the effects in different attack directions, as shown in Fig. \ref{fig:delta_SRDA}(a). Compared with Ball\'e2016, Cai2022, and Zou2022, whose performance variations are concentrated on one dimension of either distortion or bitrate, Ball\'e2018, Minnen2018, and Cheng2020 exhibit moderate adversarial robustness in both dimensions. Besides, their performance degradation directions maintain a good consistency with the attack directions. In contrast, the reconstruction quality of Song2021 and Liu2023 exhibit the largest degradation in specific directions of joint bitrate and distortion attacks. This indicates that there exits entanglement between the vulnerabilities of bitrate and distortion, which makes the joint rate-distortion attack more effective than a single dimension attack.

\subsubsection{Robustness of Submodels}
We further calculate the averaged performance variations over all the attack directions to obtain the adversarial robustness of each submodel, as shown in Fig. \ref{fig:delta_SRDA}(b). Despite sharing the same network structure, there does not exhibit strict regularity among the submodels of the same compression method. This phenomenon highlights the influence of factors other than network structure on robustness, such as the initialization of network parameters and the randomly cropped training samples.

\subsection{Performance Variations Under ARDA}
\begin{figure*}[t]
	\centering
	\includegraphics[width=0.8\linewidth]{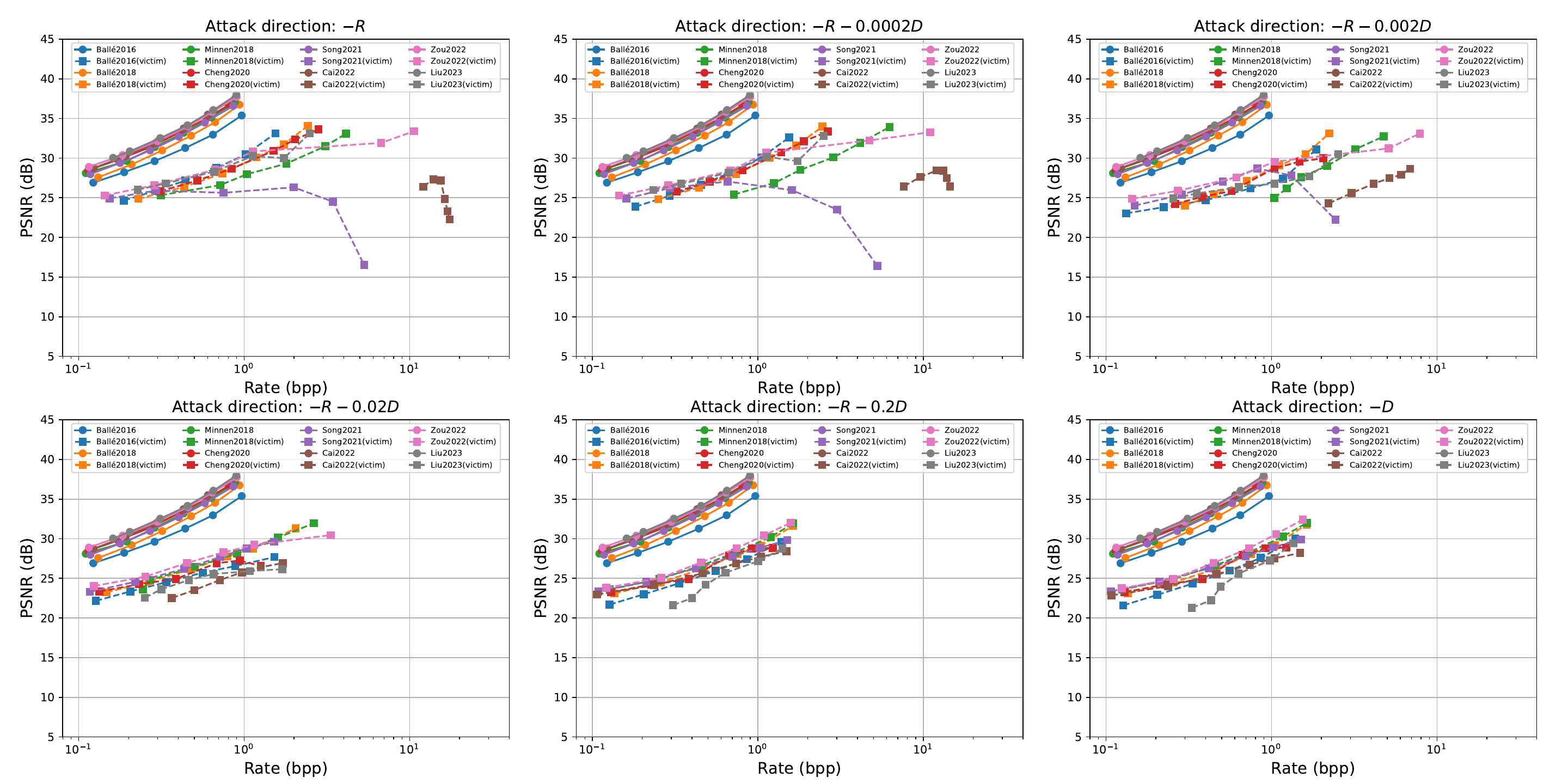}
	\caption{RD performance under six attack directions of ARDA. The performance of the same compression algorithm before and after an attack is depicted using solid and dashed lines in the same color.}
	\label{fig:rd_curves_ARDA}
\end{figure*}

Similar to the presentation of SRDA, we plot the RD performance before and after ARDA in Fig. \ref{fig:rd_curves_ARDA}. Overall, all the experimented compression methods exhibit better adversarial robustness against ARDA than SRDA. Since the attackers of ARDA try to balance the attack effects on each submodel, the RD curve of ARDA looks smoother and more regular.

\begin{table}[t]
	\centering
	\caption{Averaged performance variations $(\downarrow)$ over all submodels and attack directions. The {\color{blue}best} and {\color{red}worst} adversarial robustness among all methods are highlighted in blue and red.}
	\label{tab:overall_ARDA}
	\resizebox{0.8\linewidth}{!}{
		\begin{tabular}{l|rr|rr} \toprule
			\multicolumn{1}{c|}{\multirow{2}{*}{Methods}} & \multicolumn{4}{c}{ARDA} \\ \cline{2-5}
			& $\mu (\Delta \mathcal{R})$ & $\mu (\Delta \mathcal{D})$ & $\sigma (\Delta \mathcal{R})$ & $\sigma (\Delta \mathcal{D})$ \\ \hline
			Ball\'e2016 \cite{balle2017end} & {\color{blue}0.2113} & 119.35 & {\color{blue}0.1934} & 59.85 \\
			Ball\'e2018 \cite{ballevariational} & 0.4703 & 79.17 & 0.3685 & 46.54 \\
			Minnen2018 \cite{minnen2018joint} & 1.1586 & {\color{blue}72.35} & 0.8140 & {\color{blue}40.53} \\
			Cheng2020 \cite{Cheng_2020_CVPR} & 0.4964 & 96.02 & 0.3436 & 40.94 \\
			Song2021 \cite{song2021variable} & 0.7213 & {\color{red}194.10} & 0.7324 & {\color{red}208.74} \\
			Cai2022 \cite{cai2022high} & {\color{red}5.2461} & 137.18 & 0.9778 & 58.29 \\
			Zou2022 \cite{zou2022devil} & 1.4290 & 75.74 & {\color{red}1.8928} & 44.20 \\
			Liu2023 \cite{liu2023tcm} & 0.4598 & 140.42 & 0.3537 & 61.99 \\ \bottomrule
		\end{tabular}
	}
\end{table}

\begin{figure}
	\centering
	\includegraphics[width=\linewidth]{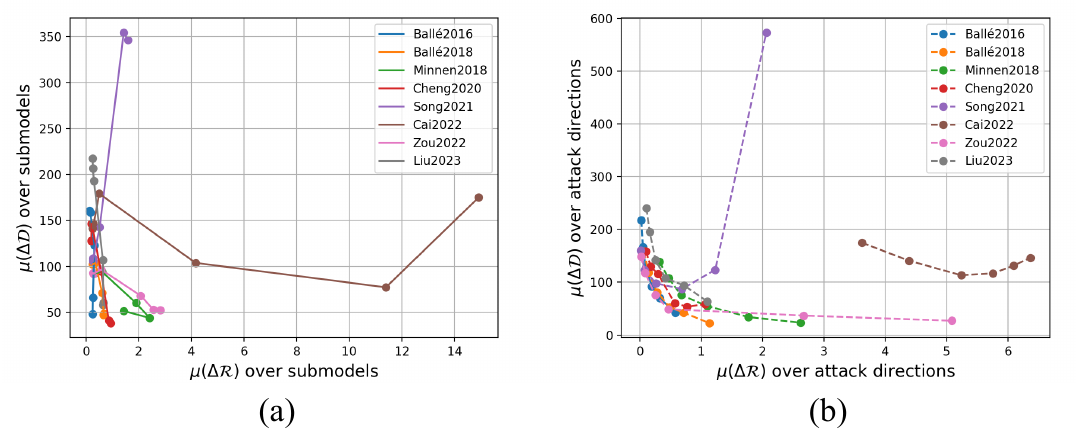}
	\caption{(a) Averaged performance variations over all submodels and (b) averaged performance variations over all attack directions of ARDA.}
	\label{fig:delta_ARDA}
\end{figure}

\subsubsection{Overall Robustness}
We calculate the mean and standard deviation of performance variations over all the submodels and attack directions under ARDA, as shown in Table \ref{tab:overall_ARDA}. Compared to SRDA, all the algorithms demonstrate notably better adversarial robustness against ARDA. This is reasonable considering that the restricted disturbance is distributed to attack submodels at all bitrates. Among them, variable-rate methods Song2021 and Cai2022 exhibit more significant performance variations than seperately trained methods. This indicates that adversarial samples are more common across their submodels.

\subsubsection{Robustness in Different Attack Directions}
Averaged performance variations $\mu(\Delta \mathcal{R})$ and $\mu(\Delta \mathcal{D})$ calculated over submodels under ARDA are shown in Fig. \ref{fig:delta_ARDA}. Performance variations in each attack direction are all significantly alleviated. Significant inferior adversarial robustness of Song2021 and Cai2022 mainly appears in the attack directions dominated by bitrate attack. This may be related to the difference between the action pattern of bitrate attack and distortion attack, which we will examine with local performance variation in sec. \ref{sec:local}.

\subsubsection{Robustness of Submodels}
Averaged performance variations $\mu(\Delta \mathcal{R})$ and $\mu(\Delta \mathcal{D})$ calculated over all the attack directions under ARDA are shown in Fig. \ref{fig:delta_ARDA}. It is observed that inferior adversarial robustness against ARDA of Song2021 is mainly manifested in the high bitrate, while it of Cai2022 is significant at all bitrates.

\subsection{Local Performance Variations}
\label{sec:local}
\begin{figure}[t]
	\centering
	\includegraphics[width=0.9\linewidth]{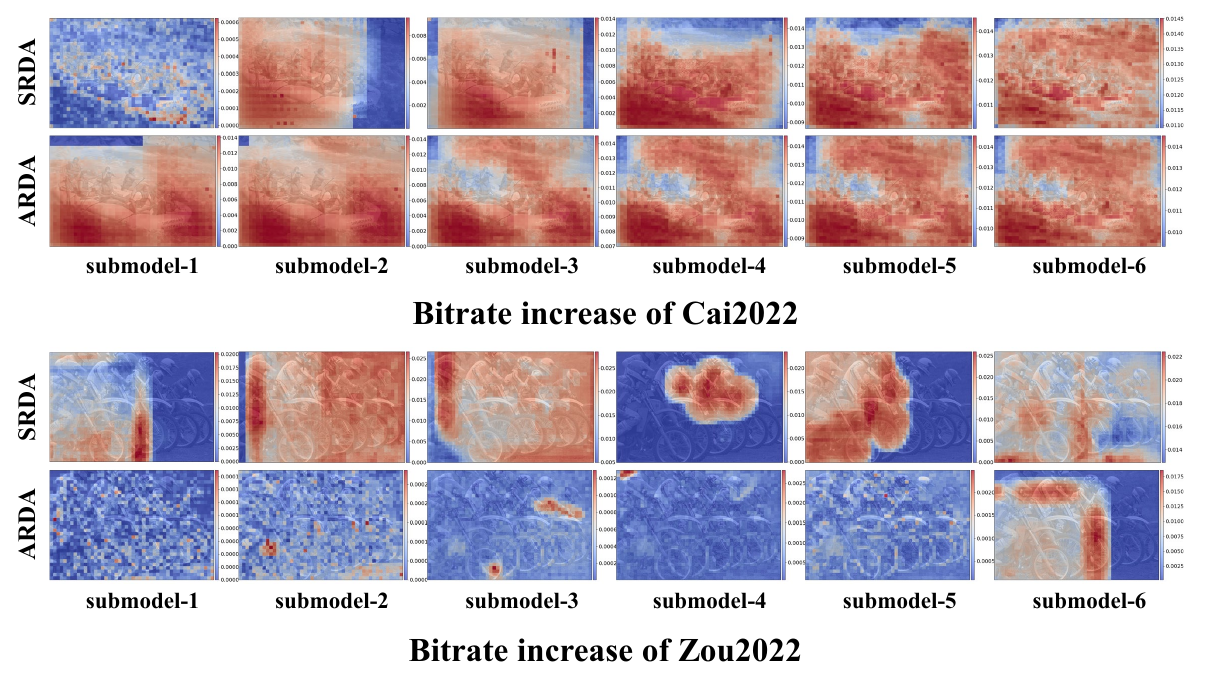}
	\caption{Patch-wise bitrate variations of Cai2022 and Zou2022, under SRDA and ARDA from the direction of bitrate attack.}
	\label{fig:local_attack_mask_rate}
\end{figure}

\begin{figure}[t]
	\centering
	\includegraphics[width=0.9\linewidth]{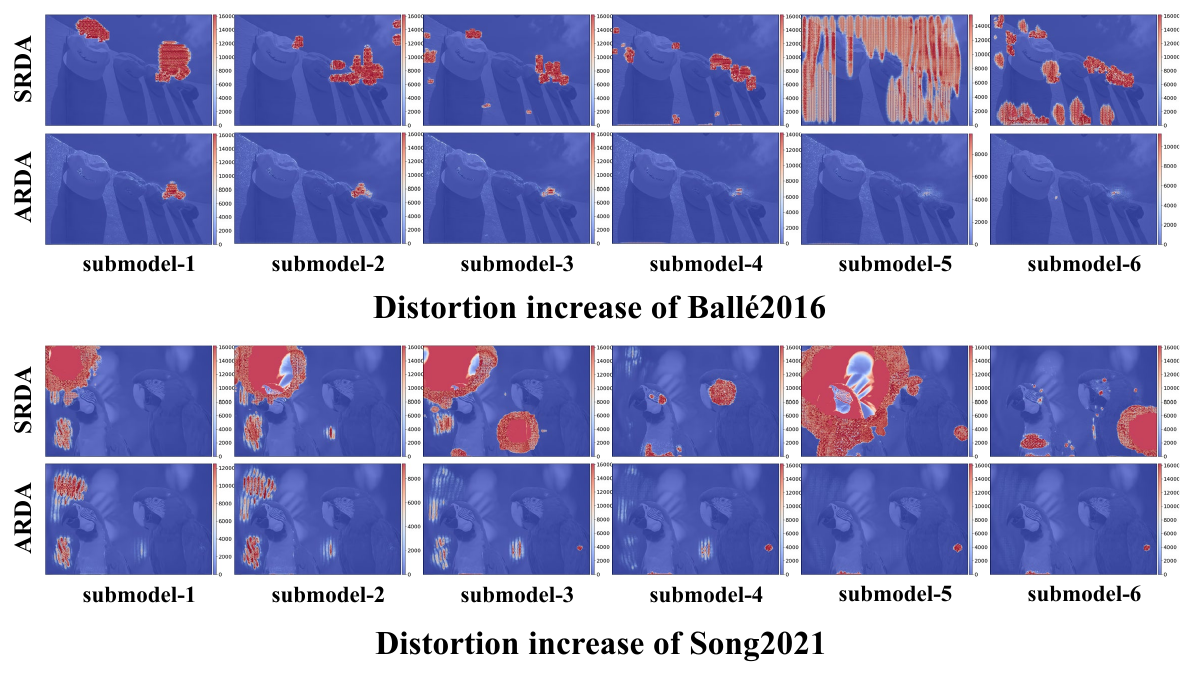}
	\caption{Pixel-wise distortion variations of Ball\'e2016 and Song2021, under SRDA and ARDA from the direction of distortion attack.}
	\label{fig:local_attack_mask_distortion}
\end{figure}

Through the local metrics defined in Eq. (\ref{eq:local}), the patch-wise bitrate variations and the pixel-wise distortion variations of four representative LIC algorithms are visualized in Fig. \ref{fig:local_attack_mask_rate} and Fig. \ref{fig:local_attack_mask_distortion}. From the visualization results, we can derive the following findings:

\subsubsection{Bitrate Attack vs. Distortion Attack} The spatial scope of bitrate attacks is extensive, resulting in varying degrees of local performance. In contrast, the spatial scope of distortion attacks is limited, but the impact on the affected area is typically significant.


\subsubsection{SRDA vs. ARDA} Distortion attack of ARDA mainly targets the intersection of vulnerable areas to SRDA in all submodels, and tends to preserve the degree of local performance variations in attacked areas. Conversely, bitrate attack of ARDA has a low correlation with the spatial position under SRDA, and primarily affects the degree of local performance variations. This difference explains why the bitrate attack of ARDA demonstrates more powerful than the distortion attack.

\subsubsection{Separately Trained vs. Variable-rate} Compared with the separately trained compression algorithms (Ball\'e2016 and Zou2022), the vulnerabilities between submodels of each variable-rate algorithm (Song2021 and Cai2022) have greater similarity. Specifically, they have larger intersection of vulnerable areas under distortion attack, and their performance variations under ARDA in the direction of bitrate attack are more significant.

\subsubsection{Optimization Instability} Variable-rate compression algorithms (Song2021 and Cai2022) are less stable than separately trained methods (Ball\'e2016 and Zou2022) in gradient-based attacks. Specifically, invalid features that exceed the maximum value of the data type often appear in Song2021. To enable the performance assessment, we replace invalid image values with $255$. This leads to the abnormal blue areas within severely distorted red areas under SRDA in Fig. \ref{fig:local_attack_mask_distortion}. In contrast, ARDA achieves better stability than SRDA by handling the gradients of all the submodels simultaneously. Consequently, ARDA demonstrates superior attack effectiveness compared to SRDA on the low bitrate submodels of Cai2022.

\begin{figure*}[t]
	\centering
	\includegraphics[width=\linewidth]{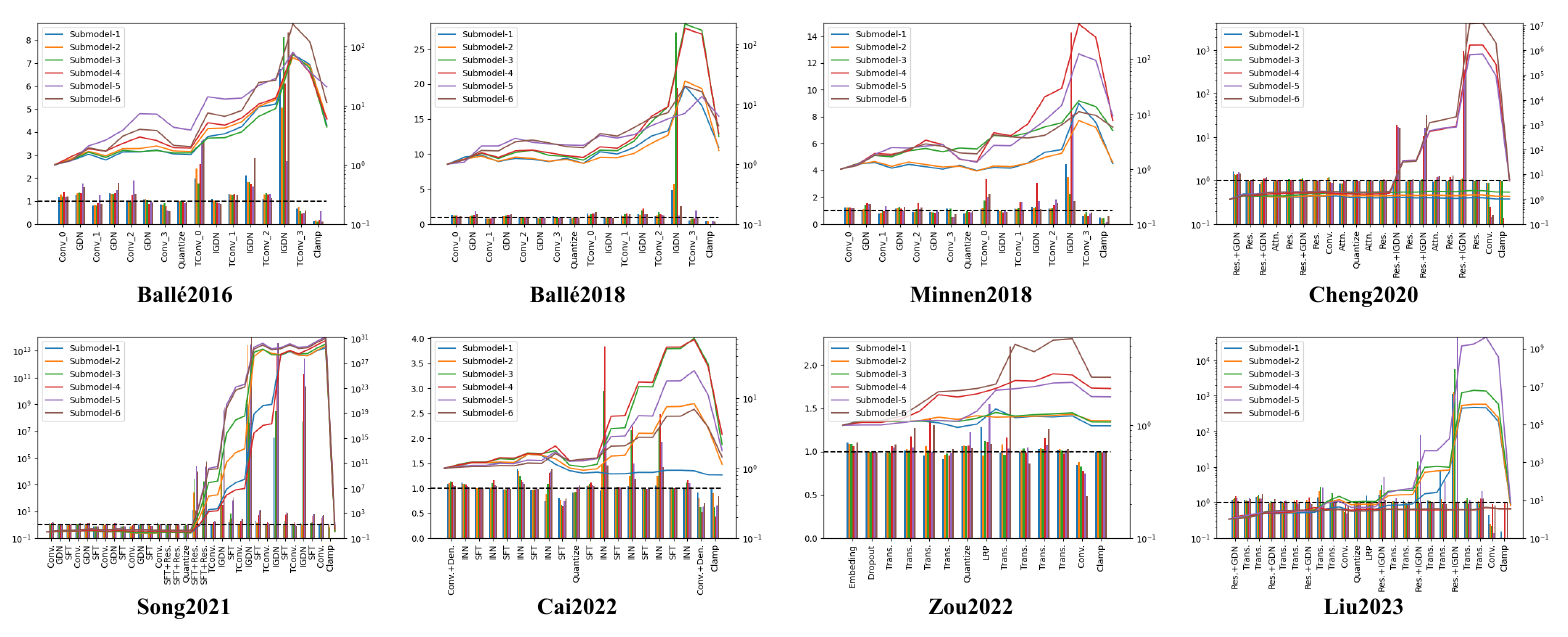}
	\caption{LDMR and CDMR of LIC algorithms under SRDA in the direction of distortion attack.}
	\label{fig:layer_response}
\end{figure*}
\begin{table}[]
	\centering
	\caption{Entropy causal intervention on LIC algorithms under SRDA in the direction of bitrate attack.}
	\label{tab:causal}
	\resizebox{\linewidth}{!}{
		\begin{tabular}{c|c|c|c|c|c|c} \toprule
			& Operation & $\Delta$Mean & Scale & Bitrate($\hat{z}$) & Bitrate($\hat{y}$) & Bitrate \\ \hline
			\multirow{2}{*}{Ball\'e2016 \cite{balle2017end}} & - & - & - & - & 1.0749 & 1.0749 \\
			& $\textbf{do}(y_s)$ & - & - & - & 0.4417 & \textbf{0.4417} \\ \hline
			\multirow{4}{*}{Ball\'e2018 \cite{ballevariational}} & - & 0.4597 & 1.5963 & 0.0899 & 3.2677 & 3.3576 \\
			& $\textbf{do}(y_h)$ & 0.4597 & 0.4497 & 0.0101 & 1.7992 & 1.8093 \\
			& $\textbf{do}(y_s)$ & 0.3089 & 1.5963 & 0.0899 & 1.6050 & \textbf{1.6949} \\ \hline
			\multirow{4}{*}{Minnen2018 \cite{minnen2018joint}} & - & 36.1617 & 19.6650 & 0.2833 & 13.3850 & 13.6683 \\
			& $\textbf{do}(y_h)$ & 0.2678 & 0.3474 & 0.0057 & 0.8348 & \textbf{0.8405} \\
			& $\textbf{do}(y_c)$ & 36.1581 & 19.6413 & 0.2833 & 13.3928 & 13.6760 \\
			& $\textbf{do}(y_s)$ & 36.1478 & 19.6650 & 0.2833 & 13.2247 & 13.5080 \\ \hline
			\multirow{4}{*}{Cheng2020 \cite{Cheng_2020_CVPR}} & - & 1.9817 & 1.7778 & 0.0443 & 3.4236 & 3.4680 \\
			& $\textbf{do}(y_h)$ & 0.4694 & 0.6504 & 0.0030 & 0.7696 & \textbf{0.7726} \\
			& $\textbf{do}(y_c)$ & 1.9653 & 1.6953 & 0.0443 & 3.5287 & 3.5730 \\
			& $\textbf{do}(y_s)$ & 1.8588 & 1.7778 & 0.0443 & 2.4796 & 2.5240 \\ \hline
			\multirow{3}{*}{Song2021 \cite{song2021variable}} & - & 3678.7946 & 4.6112 & 0.3185 & 3.2147 & 3.5332 \\
			& $\textbf{do}(y_h)$ & 0.4860 & 0.4119 & 0.0158 & 1.9108 & \textbf{1.9266} \\
			& $\textbf{do}(y_s)$ & 3678.6414 & 4.6112 & 0.3185 & 2.8378 & 3.1562 \\ \hline
			\multirow{4}{*}{Cai2022 \cite{cai2022high}} & - & 48734292.3375 & 205.4621 & 0.6646 & 15.9643 & 16.6289 \\
			& $\textbf{do}(y_h)$ & 0.4047 & 0.5395 & 0.0017 & 0.6732 & \textbf{0.6748} \\
			& $\textbf{do}(y_c)$ & 48734292.2454 & 205.4483 & 0.6646 & 16.0209 & 16.6855 \\
			& $\textbf{do}(y_s)$ & 48734292.1880 & 205.4621 & 0.6646 & 15.9372 & 16.6019 \\ \hline
			\multirow{4}{*}{Zou2022 \cite{zou2022devil}} & - & 6992.7007 & 52.4677 & 0.2423 & 19.7588 & 20.0011 \\
			& $\textbf{do}(y_h)$ & 181.3277 & 4.6411 & 0.0064 & 3.5461 & \textbf{3.5525} \\
			& $\textbf{do}(y_c)$ & 6837.7606 & 50.2262 & 0.2423 & 17.8135 & 18.0558 \\
			& $\textbf{do}(y_s)$ & 6992.6856 & 52.4677 & 0.2423 & 19.6744 & 19.9167 \\ \hline
			\multirow{4}{*}{Liu2023 \cite{liu2023tcm}} & - & 11.1363 & 0.7617 & 0.0363 & 7.6867 & 7.7230 \\
			& $\textbf{do}(y_h)$  & 0.5709 & 0.3620 & 0.0193 & 3.2734 & \textbf{3.2927} \\
			& $\textbf{do}(y_c)$  & 11.0458 & 0.7318 & 0.0363 & 7.1858 & 7.2221 \\
			& $\textbf{do}(y_s)$  & 10.9937 & 0.7617 & 0.0363 & 6.6052 & 6.5415 \\ \bottomrule
		\end{tabular}
	}
\end{table}

\subsection{Origin of Vulnerability}
To understand the origin of vulnerability in compression models, we conduct experiments with analytical tools defined in Sec. \ref{sec:tools}. Experimental results attribute the vulnerability mainly to the \textit{hyperprior} in entropy model and the \textit{IGDN} layer in transform coding. However, both of them are important components in predominant LIC algorithms. This implies that rate-distortion vulnerability widely exists in modern LIC algorithms.

\subsubsection{Vulnerability Under Bitrate Attack}
By controlling the malicious perturbations of each branch with \textit{Entropy Causal Intervention}, we present the variations of the dependent variables in Table \ref{tab:causal}. As seen from the table, replace malicious features in \textit{hyperprior} with benign features can lead to the most significant bitrate reduction on most methods. This shows that the vulnerability under bitrate attack mainly comes from the \textit{hyperprior} branch.

\subsubsection{Vulnerability Under Distortion Attack}
Through \textit{Layer-wise Distance Magnify Ratio}, we present the magnify effect of the network layers on malicious perturbations in Fig. \ref{fig:layer_response}. Compared to LDMR values which are around $1$ in Fig. \ref{fig:LDMR}, LDMR values of some network layers in Fig. \ref{fig:layer_response} have significantly increased. Through the comparison between LIC algorithms, we find that the \textit{IGDN} layer has the most significant magnification effect. Besides, compression methods Cai2022 and Zou2022 have no \textit{IGDN} layers, and they demonstrate the best distortion robustness under SRDA, which also verifies the magnification property of \textit{IGDN} on malicious noise. This suggests the central role of \textit{IGDN} in distortion vulnerability.

\subsection{Effectiveness of Defense Techniques}
\subsubsection{Adversarial Training}
\begin{figure}
	\centering
	\includegraphics[width=0.75\linewidth]{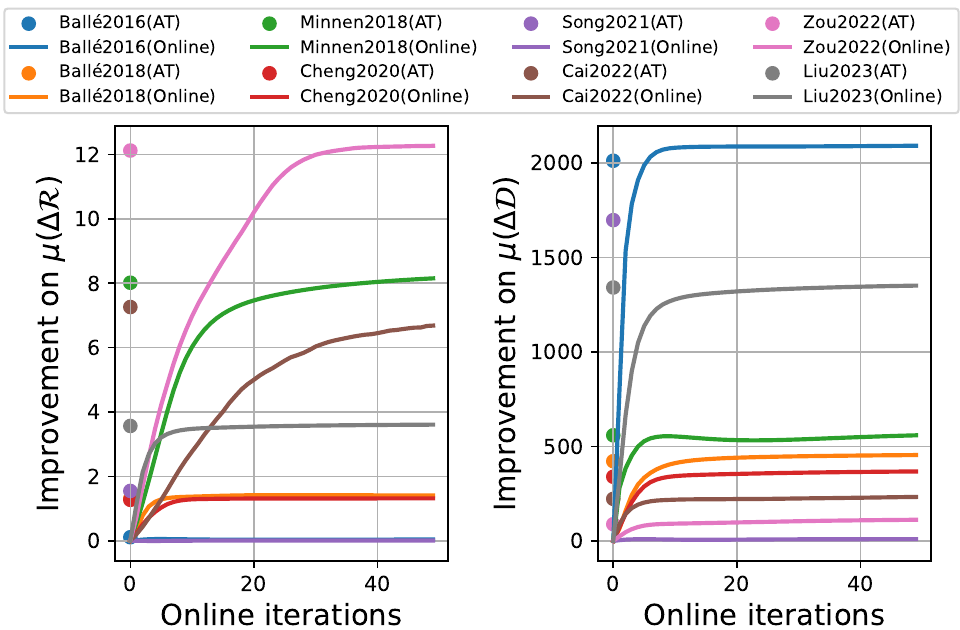}
	\caption{Improvements of adversarial training (AT) and online upadating (online) on averaged performance variations.}
	\label{fig:defense}
\end{figure}
\begin{table}
	\centering
	\caption{Averaged performance variations $(\downarrow)$ over all submodels and attack directions after adversarial training on SRA, SDA, and SRDA. The \textbf{worst} defense effect among three attack methods is highlighted in red.}
	\label{tab:defense}
	\resizebox{\linewidth}{!}{
		\begin{tabular}{l|cc|cc|cc} \toprule
			\multicolumn{1}{c|}{\multirow{2}{*}{Methods}} & \multicolumn{2}{c|}{SRA} & \multicolumn{2}{c|}{SDA} & \multicolumn{2}{c}{SRDA} \\ \cline{2-7}
			& \multicolumn{1}{c}{$\mu (\Delta \mathcal{R})$} & \multicolumn{1}{c|}{$\mu (\Delta \mathcal{D})$} & \multicolumn{1}{c}{$\mu (\Delta \mathcal{R})$} & \multicolumn{1}{c|}{$\mu (\Delta \mathcal{D})$} & \multicolumn{1}{c}{$\mu (\Delta \mathcal{R})$} & \multicolumn{1}{c}{$\mu (\Delta \mathcal{D})$} \\ \hline
			Ball\'e2016 \cite{balle2017end} &0.1817&\textbf{518.39}&\textbf{0.1934}&129.75&0.1864&147.48 \\
			Ball\'e2018 \cite{ballevariational} &0.3122&\textbf{296.39}&\textbf{1.0912}&86.67&0.3495&100.11 \\
			Minnen2018 \cite{minnen2018joint} &0.3841&\textbf{158.47}&\textbf{3.7523}&76.29&0.5095&90.36 \\
			Cheng2020 \cite{Cheng_2020_CVPR} &0.2882&\textbf{126.35}&\textbf{1.1138}&88.38&0.2580&94.88 \\
			Song2021 \cite{song2021variable} &0.4342&608.35&\textbf{1.0251}&\textbf{1064.87}&0.2876&99.42 \\
			Cai2022 \cite{cai2022high} &0.2711&143.75&\textbf{6.8692}&\textbf{257.62}&0.3786&88.35 \\
			Zou2022 \cite{zou2022devil} &0.4064&\textbf{130.15}&\textbf{2.6190}&92.47&0.3798&91.75 \\
			Liu2023 \cite{liu2023tcm} &0.3868&\textbf{132.19}&\textbf{0.7491}&94.32&0.3692&86.41 \\ \bottomrule
		\end{tabular}
	}
\end{table}
\begin{figure}[b]
	\centering
	\includegraphics[width=0.7\linewidth]{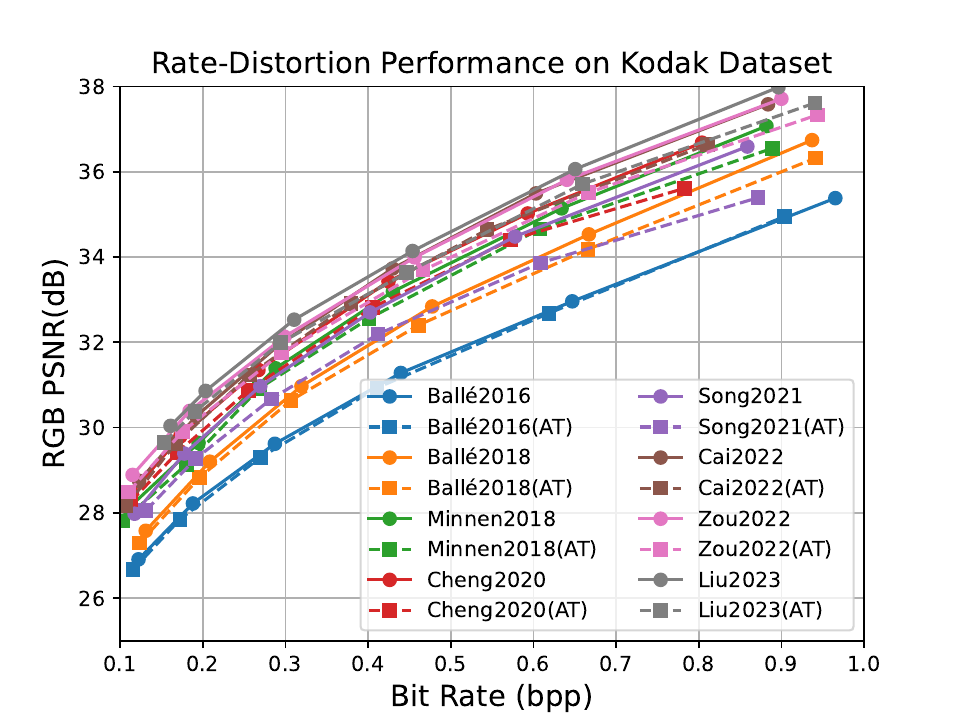}
	\caption{RD performance before and after AT on Kodak dataset.}
	\label{fig:rd_Kodak}
\end{figure}
To evaluate the defense efficacy of adversarial training, we finetune compression models on malicious samples genearted by SRDA with all attack directions. Then we assess the adversarial robustness of adversarial trained models against SRDA. It is seen from Fig. \ref{fig:defense} that AT can significantly improve the adversarial robustness of compression models.

To reveal the dependence of adversarial training on malicious samples, we further compare the adversarial robustness of compression finetuned on samples generated by bitrate attack (SRA) and distortion attack (SDA). Comparison results are shown in Table \ref{tab:defense}. It is evident that training on malicious samples generated by attacking on a single dimension of bitrate or distortion is insufficient. This highlights the importance of the proposed joint rate-distortion attack.

In addition to the dependency on the coverage of adversarial samples, another major disadvantage of AT is that it changes the RD performance of the original model. Fig. \ref{fig:rd_Kodak} presents the RD curves before and after AT. It can be observed that AT brings various degree of performance degradation to all compression algorithms.

\subsubsection{Online Updating}
We also present the improvement on adversarial robustness of online updating along with online iterations in Fig. \ref{fig:defense}. Compared with AT, online updating demonstrates comparable or slightly better performance on seperately trained models. Due to the optimization instability of Song2021 and Cai2022, it is difficult to fully use the ability of online updating. However, online updating does not change the parameters of compression models, so that the RD performance of compression methods can be guaranteed at the same time of maintaining good adversarial robustness.

\begin{table}[t]
	\centering
	\caption{Entropy causal intervention by simultaneously applying \textbf{do}($y_c$) and \textbf{do}($y_s$). The \textbf{minimum} bitrate among three defense strategies are highlighted in bold.}
	\label{tab:causal_defense}
	\resizebox{0.7\linewidth}{!}{
		\begin{tabular}{c|c|c|c} \toprule
			& W/o defense & AT & Online \\ \hline
			Ball\'e2018 \cite{ballevariational} & 1.6949 & 0.7961 & \textbf{0.6407} \\ 
			Minnen2018 \cite{minnen2018joint} & 13.5138 & 2.6528 & \textbf{0.4733} \\ 
			Cheng2020 \cite{Cheng_2020_CVPR} & 2.5051 & 0.4475 & \textbf{0.4275} \\ 
			Song2021 \cite{song2021variable} & 3.1562 & \textbf{0.9193} & 3.1403 \\ 
			Cai2022 \cite{cai2022high} & 16.5860 & \textbf{1.4832} & 2.9158 \\ 
			Zou2022 \cite{zou2022devil} & 17.8228 & 0.5046 & \textbf{0.4944} \\ 
			Liu2023 \cite{liu2023tcm} & 5.7021 & 0.6001 & \textbf{0.5692} \\ \bottomrule
		\end{tabular}
	}
\end{table}
\begin{figure}[t]
	\centering
	\includegraphics[width=\linewidth]{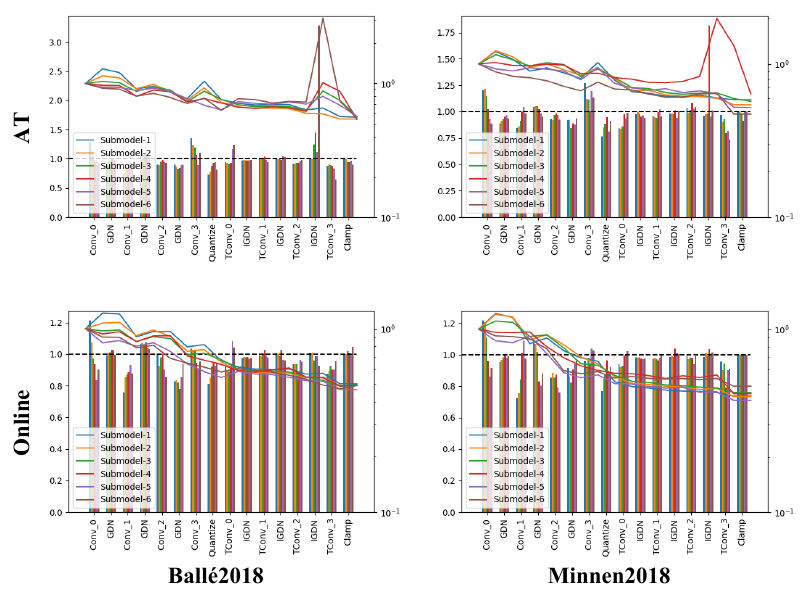}
	\caption{LDMR and CDMR after applying adversarial training and online updating.}
	\label{fig:layer_response_defense}
\end{figure}

\subsubsection{Handling of Vulnerabilities}
We further compare the ways of two defense techniques in handling the major vulnerabilities of the compression model. By simultaneously applying do-operators on $y_c$ and $y_s$, we can derive the influence of hyperprior on the final bitrate, as shown in Table \ref{tab:causal_defense}. It can be observed that both AT and online updating can obtain better adversarial robustness in hyperprior entropy model. By comparison, online updating demonstrates slightly better performance on seperately trained model, but is inferior in variable-rate methods due to the unstable optimization procedure.

Take Ball\'e2018 and Minnen2018 as examples, we compare the LDMR of AT and online updating in Fig. \ref{fig:layer_response_defense}. It is evident that both AT and online updating can suppress the LDMR to a reasonable value of around $1$. However, \textit{IGDN} after adversarial training still exists small abnormal magnification ratio. In contrast, the magnification is carefully suppressed in online updating, implying a better ability in handling malicious disturbances.

\section{Conclusion}
In this paper, two rate-distortion attack paradigms simulating real-world attacks are proposed for learning-based image compression algorithms. Additionally, two analytical tools are introduced to identify the origin of vulnerabilities. On these basis, we have conducted extensive experiments on eight predominant LIC algorighms, the main findings can be summarized as follows:
\begin{itemize}
	\item [a)] All LIC algorithms exhibit significant performance degradation under joint rate-distortion attacks. However, they are more robust against ARDA than SRDA.
	\item [b)] Variable-rate methods show inferior robustness against ARDA compared to separately trained methods, especially in terms of bitrate.
	\item [c)] The \textit{hyperprior} significantly increases the bitrate and \textit{IGDN} significantly amplifies input perturbations when under attack.
	\item [d)] With stable optimization and increased iterations, online updating can achieve comparable performance to adversarial training without compromising performance on clean images.
\end{itemize}


\bibliographystyle{IEEEtran}
\bibliography{main}

\end{document}